\documentclass[conference]{IEEEtran}
\usepackage{url}
\usepackage[utf8]{inputenc}
\usepackage{xcolor}
\usepackage{amsmath}
\usepackage{amssymb}
\usepackage{algorithm}
\usepackage{algorithmic}

\usepackage[acronyms,nonumberlist,nopostdot,nomain,nogroupskip]{glossaries}
\usepackage{tablefootnote}
\usepackage{booktabs}
\usepackage{tabularx}
\usepackage{tikz}
\usepackage{pgfplots}
\pgfplotsset{compat=newest}
\pgfplotsset{plot coordinates/math parser=false}
\usetikzlibrary{plotmarks,patterns,decorations.pathreplacing,backgrounds,calc,arrows,arrows.meta,spy,matrix,backgrounds}
\usepgfplotslibrary{patchplots,groupplots}
\usepackage{tikzscale}
\usepackage{hyperref}

\usepackage{multirow}

\usepackage[noadjust]{cite}

\usepackage{subfigure}

\usepackage[outdir=./]{epstopdf}
\graphicspath{{./figures/}}
\DeclareGraphicsExtensions{.eps,.ps,.png}

\usepackage{mathtools}

\usepackage{dblfloatfix}    
\usepackage{colortbl}

\newacronym{quic}{QUIC}{Quick UDP Internet Connections}
\newacronym{3gpp}{3GPP}{3rd Generation Partnership Project}
\newacronym{adc}{ADC}{Analog to Digital Converter}
\newacronym{5g}{5G}{5th generation}
\newacronym{aimd}{AIMD}{Additive Increase Multiplicative Decrease}
\newacronym{am}{AM}{Acknowledged Mode}
\newacronym{amc}{AMC}{Adaptive Modulation and Coding}
\newacronym{aqm}{AQM}{Active Queue Management}
\newacronym{awgn}{AGWN}{Additive White Gaussian Noise}
\newacronym{balia}{BALIA}{Balanced Link Adaptation}
\newacronym{bdp}{BDP}{Bandwidth-Delay Product}
\newacronym{bf}{BF}{Beamforming}
\newacronym{hbf}{HBF}{Hybrid Beamforming}
\newacronym{abf}{ABF}{Analog Beamforming}
\newacronym{cc}{CC}{Congestion Control}
\newacronym{cdf}{CDF}{Cumulative Distribution Function}
\newacronym{pdf}{PDF}{Probability Density Function}
\newacronym{cn}{CN}{Core Network}
\newacronym{cqi}{CQI}{Channel Quality Information}
\newacronym{cp}{CP}{Control Plane}
\newacronym{csirs}{CSI-RS}{Channel State Information - Reference Signal}
\newacronym{dc}{DC}{Dual Connectivity}
\newacronym{dce}{DCE}{Direct Code Execution}
\newacronym{dci}{DCI}{Downlink Control Information}
\newacronym{dl}{DL}{Downlink}
\newacronym{dmr}{DMR}{Deadline Miss Ratio}
\newacronym{dmrs}{DMRS}{DeModulation Reference Signal}
\newacronym{e2e}{E2E}{End-to-End}
\newacronym{ecn}{ECN}{Explicit Congestion Notification}
\newacronym{edf}{EDF}{Earliest Deadline First}
\newacronym{enb}{eNB}{evolved Node Base}
\newacronym{epc}{EPC}{Evolved Packet Core}
\newacronym{es}{ES}{Edge Server}
\newacronym{fdma}{FDMA}{Frequency Division Multiple Access}
\newacronym{fdd}{FDD}{Frequency Division Duplexing}
\newacronym[firstplural=Radio Access Technologies (RATs)]{rat}{RAT}{Radio Access Technology}
\newacronym{fs}{FS}{Fast Switching}
\newacronym{ftp}{FTP}{File Transfer Protocol}
\newacronym{gnb}{gNB}{Next Generation Node Base}
\newacronym{harq}{HARQ}{Hybrid Automatic Repeat reQuest}
\newacronym{hetnet}{HetNet}{Heterogeneous Network}
\newacronym{hh}{HH}{Hard Handover}
\newacronym{hol}{HOL}{Head-of-Line}
\newacronym{ia}{IA}{Initial Access}
\newacronym{imt}{IMT}{International Mobile Telecommunication}
\newacronym{iot}{IoT}{Internet of Things}
\newacronym{los}{LOS}{Line of Sight}
\newacronym{lte}{LTE}{Long Term Evolution}
\newacronym{m2m}{M2M}{Machine to Machine}
\newacronym{mac}{MAC}{Medium Access Control}
\newacronym{mc}{MC}{Multi-Connectivity}
\newacronym{mcs}{MCS}{Modulation and Coding Scheme}
\newacronym{mec}{MEC}{Mobile Edge Cloud}
\newacronym{mi}{MI}{Mutual Information}
\newacronym{mimo}{MIMO}{Multiple-Input Multiple-Output}
\newacronym{mumimo}{MU-MIMO}{Multi-User Multiple-Input Multiple-Output}
\newacronym{mmwave}{mmWave}{millimeter wave}
\newacronym{mr}{MR}{Maximum Rate}
\newacronym{mss}{MSS}{Maximum Segment Size}
\newacronym{mtd}{MTD}{Machine-Type Device}
\newacronym{mtu}{MTU}{Maximum Transmission Unit}
\newacronym{nfv}{NFV}{Network Function Virtualization}
\newacronym{nlos}{NLOS}{Non Line of Sight}
\newacronym{nr}{NR}{New Radio}
\newacronym{ofdm}{OFDM}{Orthogonal Frequency Division Multiplexing}
\newacronym{ofdma}{OFDMA}{Orthogonal Frequency Division Multiple Access}
\newacronym{pdcch}{PDCCH}{Physical Downlink Control Channel}
\newacronym{pucch}{PUCCH}{Physical Uplink Control Channel}
\newacronym{pdcp}{PDCP}{Packet Data Convergence Protocol}
\newacronym{pdsch}{PDSCH}{Physical Downlink Shared Channel}
\newacronym{pdu}{PDU}{Packet Data Unit}
\newacronym{pf}{PF}{Proportional Fair}
\newacronym{pgw}{PGW}{Packet Gateway}
\newacronym{phy}{PHY}{Physical}
\newacronym{pbch}{PBCH}{Physical Broadcast Channel}
\newacronym[plural=\gls{mme}s,firstplural=Mobility Management Entities (MMEs)]{mme}{MME}{Mobility Management Entity}
\newacronym{prb}{PRB}{Physical Resource Block}
\newacronym{pss}{PSS}{Primary Synchronization Signal}
\newacronym{pusch}{PUSCH}{Physical Uplink Shared Channel}
\newacronym{rach}{RACH}{Random Access Channel}
\newacronym{ran}{RAN}{Radio Access Network}
\newacronym{red}{RED}{Random Early Detection}
\newacronym{rf}{RF}{Radio Frequency}
\newacronym{rlc}{RLC}{Radio Link Control}
\newacronym{rlf}{RLF}{Radio Link Failure}
\newacronym{rrc}{RRC}{Radio Resource Control}
\newacronym{rrm}{RRM}{Radio Resource Management}
\newacronym{rr}{RR}{Round Robin}
\newacronym{rs}{RS}{Remote Server}
\newacronym{rsrp}{RSRP}{Reference Signal Received Power}
\newacronym{rss}{RSS}{Received Signal Strength}
\newacronym{rtt}{RTT}{Round Trip Time}
\newacronym{rw}{RW}{Receive Window}
\newacronym{rx}{RX}{Receiver}
\newacronym{sa}{SA}{standalone}
\newacronym{sack}{SACK}{Selective Acknowledgment}
\newacronym{sap}{SAP}{Service Access Point}
\newacronym{sch}{SCH}{Secondary Cell Handover}
\newacronym{scoot}{SCOOT}{Split Cycle Offset Optimization Technique}
\newacronym{sdma}{SDMA}{Spatial Division Multiple Access}
\newacronym{sinr}{SINR}{Signal to Interference plus Noise Ratio}
\newacronym{sm}{SM}{Saturation Mode}
\newacronym{snr}{SNR}{Signal to Noise Ratio}
\newacronym{son}{SON}{Self-Organizing Network}
\newacronym{ss}{SS}{Synchronization Signal}
\newacronym{srs}{SRS}{Sounding Reference Signal}
\newacronym{sss}{SSS}{Secondary Synchronization Signal}
\newacronym{tb}{TB}{Transport Block}
\newacronym{tcp}{TCP}{Transmission Control Protocol}
\newacronym{tdd}{TDD}{Time Division Duplexing}
\newacronym{tdma}{TDMA}{Time Division Multiple Access}
\newacronym{tfl}{TfL}{Transport for London}
\newacronym{tm}{TM}{Transparent Mode}
\newacronym{trp}{TRP}{Transmitter Receiver Pair}
\newacronym{tti}{TTI}{Transmission Time Interval}
\newacronym{ttt}{TTT}{Time-to-Trigger}
\newacronym{tx}{TX}{Transmitter}
\newacronym{ue}{UE}{User Equipment}
\newacronym{ul}{UL}{Uplink}
\newacronym{uml}{UML}{Unified Modeling Language}
\newacronym{um}{UM}{Unacknowledged Mode}
\newacronym{utc}{UTC}{Urban Traffic Control}
\newacronym{vm}{VM}{Virtual Machine}
\newacronym{rsrq}{RSRQ}{Reference Signal Received Quality}
\newacronym{rssi}{RSSI}{Received Signal Strength Indicator}
\newacronym{crs}{CRS}{Cell Reference Signal}
\newacronym{comp}{CoMP}{Coordinated Multi-Point}
\newacronym{cran}{C-RAN}{Cloud \acrlong{ran}}
\newacronym{ca}{CA}{Carrier Aggregation}
\newacronym{cco}{CC}{Carrier Component}
\newacronym{nsa}{NSA}{Non Stand Alone}
\newacronym{embb}{eMBB}{Enhanced Mobility Broadband}
\newacronym{bsr}{BSR}{Buffer Status Report}
\newacronym{srb}{SRB}{Service Radio Bearer}
\newacronym{scm}{SCM}{Spatial Channel Model}
\newacronym{sctp}{SCTP}{Stream Control Transmission Protocol}
\newacronym{mptcp}{MPTCP}{Multi-path TCP}
\newacronym{ietf}{IETF}{Internet Engineering Task Force}
\newacronym{os}{OS}{Operating System}
\newacronym{tls}{TLS}{Transport Layer Security}
\newacronym{rfc}{RFC}{Request for Comments}
\newacronym{http}{HTTP}{HyperText Transfer Protocol}
\newacronym{nat}{NAT}{Network Address Translation}
\newacronym{api}{API}{Application Programming Interface}
\newacronym{rto}{RTO}{Retransmission Timeout}
\newacronym{psc}{PSC}{Public Safety Communication}
\newacronym{rpgm}{RPGM}{Reference Point Group Mobility}
\newacronym{ic}{IC}{Incident Command}
\newacronym{rsu}{RSU}{Road Side Unit}
\newacronym{uav}{UAV}{Unmanned Aerial Vehicle}
\newacronym{iab}{IAB}{Integrated Access and Backhaul}
\newacronym{psd}{PSD}{Power Spectral Density}
\newacronym{mpc}{MPC}{Multi Path Component}
\newacronym{rt}{RT}{Ray Tracer}
\newacronym{aoa}{AoA}{Angle of Arrival}
\newacronym{aod}{AoD}{Angle of Departure}
\newacronym{inr}{INR}{Interference to Noise Ratio}
\newacronym{qd}{QD}{Quasi Deterministic}
\newacronym{wlan}{WLAN}{Wireless Local Area Network}
\newacronym{cad}{CAD}{Computer-aided Design}
\newacronym{ap}{AP}{Access Point}
\newacronym{sta}{STA}{Station}
\newacronym{nrmse}{NRMSE}{Normalized Root Mean Square Error}
\newacronym{ut}{UT}{User Terminal}
\newacronym{bs}{BS}{Base Station}
\newacronym{mmse}{MMSE}{Minimum Mean Squared Error}
\newacronym{gbf}{GBF}{Geometric BeamForming}
\newacronym{cbf}{CBF}{Codebook BeamForming}
\newacronym{fmbf}{FMBF}{Frequency-Flat MMSE BeamForming}
\newacronym{smbf}{SMBF}{Frequency-Selective MMSE BeamForming}
\newacronym{bler}{BLER}{Block Error Rate}
\newacronym{fft}{FFT}{Fast Fourier Transform}
\newacronym{nack}{NACK}{Negative Acknowledgment}
\newacronym{upa}{UPA}{Uniform Planar Array}
\newacronym{tmrs}{TMRS}{TDMA mmWave RR Scheduler}
\newacronym{pmrs}{PMRS}{Padded mmWave RR Scheduler}
\newacronym{amrs}{AMRS}{Asynchronous mmWave almost-RR Scheduler}
\newacronym{rb}{RB}{Resource Block}
\newacronym{udp}{UDP}{User Datagram Protocol}
\newacronym{noma}{NOMA}{Non Orthogonal Multiple Access}
\newacronym{dft}{DFT}{Discrete Fourier Transform}
\newacronym{cav}{CAV}{Connected Autonomous Vehicles}
\newacronym{fr2}{FR2}{Frequency Range 2}

\newcommand{\Hb}{\mathbf{H}}

\newcommand{\V}{\mathbf{V}}

\newcommand{\I}{\mathbf{I}}

\newcommand{\x}{\mathbf{x}}

\newcommand{\vv}{\mathbf{v}}

\newcommand{\y}{\mathbf{y}}

\newcommand{\w}{\mathbf{w}}

\newcommand{\ab}{\mathbf{a}}

\usepackage{pifont}
   \definecolor{blueH3}{rgb}{0,.5,1}
   \definecolor{blueH2}{rgb}{0,0.25,0.75}
   \definecolor{blueH1}{rgb}{0,0,0.5}   
   \definecolor{grayOldText}{rgb}{.5,.5,.5}
   \definecolor{VCobalt}{HTML}{005682}
   \definecolor{TZTeal}{HTML}{008080}
   \definecolor{KYJade}{HTML}{008151}
   \definecolor{ARust}{HTML}{a10000}
   \definecolor{FFucsia}{HTML}{7000c3}
   
\newcommand{\CASE}[1]{\STATE \textbf{case} #1\textbf{:} \begin{ALC@g}}
\newcommand{\ENDCASE}{\end{ALC@g}}

\newcommand{\DEFAULT}{\STATE \textbf{default:} \begin{ALC@g}}
\newcommand{\ENDDEFAULT}{\end{ALC@g}}
\newcommand{\DEFAULTLINE}[1]{\STATE \textbf{default:} }

\usepackage{tikz}
\usepackage{pgfplots}
\pgfplotsset{compat=newest} 
\pgfplotsset{plot coordinates/math parser=false} 
\newlength\fheight
\newlength\fwidth
\usetikzlibrary{plotmarks,patterns,decorations.pathreplacing,backgrounds,calc,arrows,arrows.meta,spy,matrix}
\usepgfplotslibrary{patchplots,groupplots}
\usepackage{tikzscale}

\tikzstyle{startstop} = [rectangle, rounded corners, minimum width=2cm, minimum height=0.5cm,text centered, draw=black]
\tikzstyle{io} = [trapezium, trapezium left angle=70, trapezium right angle=110, minimum width=3cm, minimum height=1cm, text centered, draw=black]
\tikzstyle{process} = [rectangle, minimum width=2cm, minimum height=0.5cm, text centered, draw=black, align=center]
\tikzstyle{decision} = [ellipse, minimum width=2cm, minimum height=1cm, text centered, draw=black]
\tikzstyle{arrow} = [thick,<->,>=stealth]
\tikzstyle{line} = [thick,>=stealth]
\tikzstyle{darrow} = [thick,<->,>=stealth,dashed]
\tikzstyle{sarrow} = [thick,->,>=stealth]
\tikzstyle{larrow} = [line width=0.1mm,dashdotted,<->,>=stealth]

\pgfplotsset{every tick label/.append style={font=\scriptsize}, 
             every axis/.append style={
             width=\fwidth, height=\fheight, at={(0\fwidth,0\fheight)}, 
             xlabel style={font=\footnotesize\color{white!15!black}},
             xmajorgrids,
             ylabel style={yshift=-0.15cm, font=\footnotesize\color{white!15!black}},
             ymajorgrids,
             legend style={font=\footnotesize\color{white!15!black}},
             /pgfplots/ybar legend/.style={/pgfplots/legend image code/.code={\draw[##1,/tikz/.cd,yshift=-0.25em](0cm,0cm) rectangle (10pt,1em);},},
             }}

\definecolor{SchoolColor}{RGB}{0.71, 0, 0.106}
\definecolor{chaptergrey}{rgb}{0.61, 0, 0.09} 
\definecolor{midgrey}{rgb}{0.4, 0.4, 0.4}
\definecolor{chaptergreen}{rgb}{0.09, 0.612, 0}
\definecolor{chapterpurple}{rgb}{0.522, 0, 0.612}
\definecolor{chapterlightgreen}{rgb}{0, 0.612, 0.522}

\makeatletter
\def\grd@save@target#1{%
  \def\grd@target{#1}}
\def\grd@save@start#1{%
  \def\grd@start{#1}}
\tikzset{
  grid with coordinates/.style={
    to path={%
      \pgfextra{%
        \edef\grd@@target{(\tikztotarget)}%
        \tikz@scan@one@point\grd@save@target\grd@@target\relax
        \edef\grd@@start{(\tikztostart)}%
        \tikz@scan@one@point\grd@save@start\grd@@start\relax
        \draw[minor help lines] (\tikztostart) grid (\tikztotarget);
        \draw[major help lines] (\tikztostart) grid (\tikztotarget);
        \grd@start
        \pgfmathsetmacro{\grd@xa}{\the\pgf@x/1cm}
        \pgfmathsetmacro{\grd@ya}{\the\pgf@y/1cm}
        \grd@target
        \pgfmathsetmacro{\grd@xb}{\the\pgf@x/1cm}
        \pgfmathsetmacro{\grd@yb}{\the\pgf@y/1cm}
        \pgfmathsetmacro{\grd@xc}{\grd@xa + \pgfkeysvalueof{/tikz/grid with coordinates/major step x}}
        \pgfmathsetmacro{\grd@yc}{\grd@ya + \pgfkeysvalueof{/tikz/grid with coordinates/major step y}}
        \foreach \x in {\grd@xa,\grd@xc,...,\grd@xb}
        \node[anchor=north] at (\x,\grd@ya) {\pgfmathprintnumber{\x}};
        \foreach \y in {\grd@ya,\grd@yc,...,\grd@yb}
        \node[anchor=east] at (\grd@xa,\y) {\pgfmathprintnumber{\y}};
      }
    }
  },
  minor help lines/.style={
    help lines,
    gray,
    line cap =round,
    xstep=\pgfkeysvalueof{/tikz/grid with coordinates/minor step x},
    ystep=\pgfkeysvalueof{/tikz/grid with coordinates/minor step y}
  },
  major help lines/.style={
    help lines,
    line cap =round,
    line width=\pgfkeysvalueof{/tikz/grid with coordinates/major line width},
    xstep=\pgfkeysvalueof{/tikz/grid with coordinates/major step x},
    ystep=\pgfkeysvalueof{/tikz/grid with coordinates/major step y}
  },
  grid with coordinates/.cd,
  minor step x/.initial=.5,
  minor step y/.initial=.2,
  major step x/.initial=1,
  major step y/.initial=1,
  major line width/.initial=1pt,
}
\makeatother

\DeclareUnicodeCharacter{2212}{-} 

\glsdisablehyper
\begin{document}

\title{Full-stack Hybrid Beamforming\\ in mmWave 5G Networks}

  \author{
  \IEEEauthorblockN{Felipe G\'omez-Cuba$^1$, Tommaso Zugno$^2$, Junseok Kim$^3$, Michele Polese$^4$, Saewoong Bahk$^5$, Michele Zorzi$^2$}
   
\IEEEauthorblockA{
    $^1$AtlantTIC, University of Vigo, Spain. Email: \texttt{gomezcuba@gts.uvigo.es}\\
    $^2$Department of Information Engineering, University of Padova, Padova, Italy.\\Email: \texttt{\{zugnotom, zorzi\}@dei.unipd.it}\\
    $^3$System LSI, Samsung Electronics, Gyeonggi-do, South Korea. Email: \texttt{junseok.kim@samsung.com}\\
    $^4$Institute for the Wireless Internet of Things, Northeastern University, Boston, MA USA.\\Email: \texttt{m.polese@northeastern.edu}\\
    $^5$Department of ECE and INMC, Seoul National University, Seoul, South Korea. Email: \texttt{sbahk@snu.ac.kr.}
    }
}

\flushbottom
\setlength{\parskip}{0ex plus0.1ex}

\maketitle
\thispagestyle{empty}

\glsunset{nr}

\begin{abstract}
This paper analyzes \gls{hbf} and \gls{mumimo} in \gls{mmwave} \gls{5g} cellular networks considering the full protocol stack with TCP/IP traffic and MAC scheduling. Prior work on \gls{hbf} and \gls{mumimo} has assumed full-buffer transmissions and studied link-level performance. We report non-trivial interactions between the \gls{hbf} technique, the front-loaded channel estimation pilot scheme in \gls{nr}, and the constraints of \gls{mumimo} scheduling. We also report that joint multi-user beamforming design is imperative, in the sense that the \gls{mumimo} system cannot be fully exploited when implemented as a mere collection of single-user analog beams working in parallel. By addressing these issues, throughput can be dramatically increased in \gls{mmwave} \gls{5g} networks by means of \gls{sdma}.
\end{abstract}


\section{Introduction}

The next generations of mobile networks will need to support a wide range of applications that demand significantly higher rates, such as video and content delivery; lower latency, such as eHealth and \glspl{cav}; and increased reliability, such as \gls{iot} in smart factories and smart cities~\cite{BocHLMP:14}.
The \gls{3gpp} specification for \gls{5g} cellular networks~\cite{38300} is positioned to address these challenges, introducing a new \gls{ran} design (i.e., 3GPP \gls{nr}). It features a flexible \gls{ofdm} frame structure with a flexible \textit{numerology}, and, for the first time in the \gls{ran}, the support for 
\gls{mmwave} communications in the $24$--$53$~GHz band, referred to as \gls{fr2}. 

\gls{nr} in the \gls{mmwave} spectrum supports much larger bandwidths with respect to legacy \gls{3gpp} \glspl{rat} in sub-6 GHz bands~\cite{RanRapE:14}, with up to 400 MHz for each carrier~\cite{38300}. The small wavelength permits to integrate a large number of antenna elements even in a small mobile device. By means of \gls{bf} techniques it is possible to concentrate the transmitted power in a single direction and to offset the higher path loss at higher carrier frequencies. Since 
the use of fully digital large antenna arrays with high-resolution \glspl{adc} consumes too much power, energy efficient \gls{bf} architectures are considered. In \gls{abf} and \gls{hbf} architectures, one or $K>1$ \gls{rf} chains with analog circuits feed an array of $N\gg K$ antennas, respectively. In \gls{abf}, a single frequency-flat analog beam can be employed at a time, whereas in hybrid \gls{bf} more sophisticated beam design is possible at the expense of some more power consumption~\cite{7400949}.  Other proposals consider the usage of low-resolution \glspl{adc} in fully-digital \gls{bf} architectures~\cite{7400949}, which we leave for consideration in future work.

\gls{hbf} is capable of using digital low-dimensional linear operations to modify the effective beams seen by different \gls{ofdm} subcarriers in a frequency-selective manner. In addition, \gls{hbf} is capable of steering multiple \gls{sdma} \textit{layers} at the same time, i.e., different signal beams delivering independent data streams. It is typical that \gls{mmwave} channels are sparse in the sense that sending multiple \gls{sdma} layers to the same user is ineffective, making \gls{mumimo} imperative to achieve spatial multiplexing gains in \glspl{mmwave}~\cite{SunRap:cm14}.

Abundant physical layer literature has studied beam design, transceiver circuits, and spatial multiplexing for \gls{mmwave}, such as \cite{7400949,SunRap:cm14,sohrabi2017hybrid,kulkarni2016comparison}. Nowadays the \gls{3gpp} \gls{nr} specifications support \gls{hbf} and \gls{mumimo} \cite{38300,8692922}. However, the state of the art currently lacks an analysis of how a \textit{physical layer} based on \gls{hbf} interacts with the \textit{full protocol stack}. 

There is rising recent interest on the evaluation of the end-to-end performance of \gls{5g} mmWave networks, with analysis, simulations, and experiments. However, most end-to-end simulation literature has focused on single-layer analog \gls{bf} \cite{mezzavilla2018end, choi20195g}, whereas most of the work considering \gls{sdma} focused on full buffer link-level evaluation \cite{7400949,SunRap:cm14,sohrabi2017hybrid,kulkarni2016comparison}.
To fill this gap, in this paper we study the integration of \gls{hbf} techniques for \gls{mumimo} \gls{sdma} into a full-stack \gls{5g} and beyond cellular network, focusing on beam design in presence of inter-beam interference and on the scheduling constraints arising in a context of multiple simultaneous transmissions. 

We extend prior models for full-stack \gls{mmwave} \gls{5g} networks by considering \gls{mumimo} transmissions. We find that the \gls{cbf} schemes that work well in the single-user case lead to high inter-beam interference in a \gls{sdma} scenario, and propose a \gls{smbf}, which performs low-dimensional linear pre-processing for \gls{hbf} beam design in order to remove this inter-beam interference. We notice a suboptimal interaction between scheduling and \gls{mumimo} beam design. This arises from the fact that the \textit{effective} channel coefficient (after beamforming is applied) needs to remain unchanged during an entire allocated period of time where a single pilot is used to estimate the \gls{ofdm} channel gains. As a consequence, to mitigate the inter-beam interference through the introduction of a \gls{smbf} scheme, all simultaneous transmissions must begin at the same time. To this aim, the scheduler needs to allocate some padding blank resources without a transmission, leading to resource waste.

Finally, we demonstrate our results in a novel \gls{hbf} extension\footnote{Available at \url{https://github.com/signetlabdei/ns3-mmwave-hbf}} for the ns-3 \gls{mmwave} module~\cite{mezzavilla2018end}. We believe this is the first open source software to support \gls{hbf} \gls{mumimo} \gls{sdma} at \glspl{mmwave} with 3GPP-like \gls{mac}, the \gls{3gpp} \gls{mmwave} channel model, and realistic TCP/IP traffic. Our results show that \gls{mumimo} \gls{sdma} relying on \gls{hbf} can increase the capacity of a single-layer \gls{mmwave} network, but \gls{hbf}-aware scheduling design is fundamental to achieve the potential gains. Additional material and results can be found in the draft available in~\cite{cuba2020hybrid}.

The rest of the paper is structured as follows. Sec.~\ref{sec:system} discusses our \gls{hbf} and scheduling model for \gls{3gpp} \gls{nr} networks. Sec.~\ref{sec:perf_eval} describes the performance evaluation results. Finally, Sec.~\ref{sec:conclusions} concludes the paper.

\section{Full-Stack Integration of \gls{hbf} for \glspl{mmwave}}
\label{sec:system} 

In the \gls{nr} frame, complex symbols are mapped in a 3-dimensional \gls{ofdm} resource grid, comprising the \gls{ofdm} symbol number in time ($n$), the \gls{ofdm} subcarrier number in frequency ($k$), and the \textit{\gls{sdma} layer} number ($\ell$) \cite{TS38211v16}. Some options of the waveform are suited for sub-$6$~GHz frequencies but not for \glspl{mmwave}, for instance, since \gls{hbf} features frequency-flat analog operations that cannot differ for different \gls{ofdm} subcarriers, \gls{ofdma} is not employed at FR2 and all subcarriers $k$ in a port-symbol pair are assigned to the same user. Thus, the scheduling reduces to a 2-dimension \gls{tdma} and \gls{sdma} grid $(n,\ell)$. Moreover, typical \gls{mmwave} \gls{mimo} channel matrices are rank deficient (i.e., the second largest eigenvalue is much smaller than the first) \cite{SunRap:cm14}. This means that assigning two or more \gls{sdma} layers to the same user is ineffective. Finally, the layers are mapped to one or more antenna ports ($p$), defined as an \gls{rf} input to the array that experiences the same frequency-flat analog circuitry, using low dimensional digital linear precoding that can be frequency-selective \cite{TS38211v16}.

The channel matrix between the \gls{bs} and each \gls{ue} $u$ is denoted by $\Hb_{u}[n,k]$ in \gls{ofdm} symbol $n$ and subcarrier $k$. In \gls{dl}, the BS selects a \gls{bf} vector for each layer and subcarrier $\vv_\ell[k]$ using a certain \gls{bf} scheme, and the UE receives with the analog \gls{bf} vector $\w_u$. Thus, the \textit{effective} scalar complex channel between the transmit port $p$ and the receive antenna port of the UE is given by
$$h_{eq}[u,\ell,n,k]=\w_u^T[k]\Hb_{u}[n,k]\vv_\ell[k],$$
while the \gls{ul} channel is computed with the transposed channel matrix and swapping transmitter and receiver beamforming vectors, resulting in the same complex scalar number. 

We assume a \gls{sinr}-based point-to-point link performance model. We compute the \gls{sinr} of each link, assuming that simultaneous links are decoded separately, and map their \gls{sinr} to the \gls{bler} of the transmission. This is a simplification of real decoding hardware that makes the simulation of a large network tractable. Real \gls{nr} demodulation and decoding may use sophisticated joint decoding such as sphere decoding \cite{fincke1985improved}, as well as LDPC and Polar channel codes \cite{TS38211v16}.

We model each \gls{ofdm} symbol independently, so for the \gls{sinr} calculation we will omit the \gls{ofdm} symbol index $n$. For multiple simultaneous \gls{dl} transmissions sent by the \gls{bs} on multiple layers, we write the \gls{dl} \gls{sinr} of user $u$ at subcarrier $k$ as a function of the effective channel gains as
\begin{equation}
\label{eq:sinrdl}
    SINR_u^{DL}[k]=\frac{ L_u|h_{eq}[u,\ell(u),k]|^2P_{sc}}{\sum_{u'\neq u} L_{u}|h_{eq}[u,\ell(u'),k]|^2P_{sc}+\Delta f N_o}
\end{equation}
where $\ell(u)$ indicates the layer assigned to the \gls{ue} $u$, $L_u$ is the pathloss of $u$, $P_{sc}$ is the transmitter power per layer and per subcarrier, $N_o$ is the noise \gls{psd} and $\Delta f$ is the inter-carrier spacing. We focus on the sum of interference over \glspl{ue} connected to the same BS, producing inter-beam interference. We assume there is no cooperation and if there are other BSs, their interference may be modeled in bulk by increasing $N_o$. We note that the inter-beam interference depends on the ``mismatched'' vectors of $u$ and $u'$. Even if this reduces the \gls{bf} gain experienced by the interference, this term may be significant, making the link \gls{sinr} much lower than the \gls{snr}.

The \gls{ul} \gls{sinr} follows the expression
\begin{equation}
\label{eq:sinrul}
SINR_u^{UL}[k]=\frac{ L_u|h_{eq}[u,\ell(u),k]|^2P_{sc}}{\sum_{u'\neq u} L_{u'}|h_{eq}[u',\ell(u),k]|^2P_{sc}+\Delta f N_o},
\end{equation}
where the interference of user $u'$ is received through the channel of user $u'$, with pathloss gain $L_{u'}$. This can make the \gls{ul} interference even more severe (e.g., if $L_{u'}\gg L_u$).

This \gls{sinr} point-to-point link model can support any arbitrary \gls{bf} vector design. In this paper we consider the following \gls{bf} techniques to achieve good link \gls{snr} or \gls{sinr} values.


\subsection{CodeBook Beamforming (CBF)}

We define a \textit{\gls{bf} codebook} $\mathcal{B}$ as a small collection of possible frequency-flat analog \gls{bf} vectors. The transmitter sends reference signals using all the vectors in the set $\mathcal{B}_T$, and the receiver tests decoding the reference signals with all vectors in $\mathcal{B}_R$. Finally, the receiver selects the best pair and feeds back the decision to the transmitter. As an approximation of a real maximum received reference signal power criterion, in our implementation we assume the ideal max-\gls{snr} criterion: 
\begin{equation}
\vv_{\ell(u)},\w_u=\arg \max_{ \vv \in\mathcal{B}_T,\w \in\mathcal{B}_R} |\w_u^T\Hb_{u}[n,k_{ref}]\vv_\ell|^2,
\end{equation}
where $k_{ref}$ is a single subcarrier index where we assume a narrowband reference signal was sent.

We denote the antenna array response as a function $\ab(\theta,\phi)$ that depends on the angles of azimuth and elevation $(\theta,\phi)$. In our simulations we adopt the \gls{upa} model with $N_1\times N_2$ antennas separated half a wavelength, where the $i$-th element of the vector is $$a_i(\theta,\phi)=e^{-j\frac{\pi}{2}\left((i \mod N_1)\sin(\theta) + \lfloor i / N_1\rfloor\sin(\phi)\right)}.$$
By generating a codebook of $N_1N_2$ vectors using $\ab()$ at the special angles $\{\theta = \sin^{-1}(\frac{2n}{N_1}-1)):n\in\{0\dots N_1-1\}\}$ and $\{\phi = \sin^{-1}(\frac{2n}{N_2}-1)):n\in\{0\dots N_2-1\}\}$, then the codebook conveniently becomes the set of columns of a $N_1N_2\times N_1N_2$ double \gls{dft} matrix. The \gls{dft} codebook facilitates implementation with analog phase-arrays or with lensed arrays, and is similar to precoding in \cite{TS38211v16}.

\subsection{Frequency-Selective MMSE BeamForming (SMBF)}

This \gls{bf} scheme introduces a low-complexity, low-dimensional, frequency-selective linear matrix mapping between layers and ports, in combination with an auxiliary analog frequency-flat \gls{cbf} underlying scheme. Let us denote the \gls{bf} vectors selected using \gls{cbf} by $\w_u^{CB}$ and $\vv_\ell^{CB}$. We assume that first the system conducts a codebook exploration as in \gls{cbf} and loads the best codebook \gls{bf} vector for each user $u$ to different antenna ports denoted $p(u)$. In addition, we assume that after the codebook exploration the \gls{bs} transmits pilot signals in all subcarriers and the receivers can report back a set of effective channel coefficients $\{\sqrt{L_u}h_{eq}^{CB}[u,p(u'),n,k]$ for all pairs $(u,p(u'))$ and subcarrier indices $k\}$. The same pilots used for data decoding are suitable, but to report these auxiliary effective channel coefficients back to the BS would incur some overhead. Nonetheless, in this paper we assume that ideal \gls{mmse} precoding is possible, and we leave overhead reduction and other precoding constraints for future work.

To simplify notation, we assume that the users are numbered sequentially $u\in\{0\dots N_u\}$ and that their assigned layer and port numbers are equally sequential $\ell(u)=p(u)=u$. Using the auxiliary scalar channel coefficients, the BS builds the \gls{mumimo} reference equivalent channel matrix 
\begin{equation}
\label{eq:equivchanmmse}
\begin{split}
&\Hb_{eq}[k]=\\
&\;\left(\begin{array}{ccc}
\sqrt{L_1}h_{eq}^{CB}[1,1,k] & \dots & \sqrt{L_1}h_{eq}^{CB}[1,N_p,k]\\
\vdots &  \ddots & \vdots\\
\sqrt{L_{N_u}}h_{eq}^{CB}[N_u,1,k]  & \dots & \sqrt{L_{N_u}}h_{eq}^{CB}[N_u,N_p,k]\\
\end{array}\right),
\end{split}
\end{equation}
where $N_p=N_u$ is the number of analog \gls{bf} ports, each associated to a single user.
Moreover since $\ell(u)=p(u)=u$, the desired channels are in the main diagonal of this matrix.

Finally, for \gls{dl}, on the receiver side, the receiving \gls{bf} vectors would remain those of \gls{cbf}, while on the transmitter side the BS designs a set of precoding matrices for each subcarrier $k$, matching layers to ports using the following \gls{mmse} \gls{dl} precoding expression:
$$\V_{\gls{mmse}}[k]=\Hb_{eq}^H[k](\Hb_{eq}[k]\Hb_{eq}^H[k]+\frac{N_o\Delta f}{P}\I)^{-1}.$$
We adopt the \gls{mmse} technique because, when the noise is weak compared to the transmitted power, then $\frac{N_o\Delta f}{P_{sc}}\to 0$, and the expression converges to the pseudoinverse (zero-forcing precoder), i.e., $\Hb_{eq}[k]\V_{\gls{mmse}}[k]=\I$, thus suppressing the interference. In addition when the noise is strong, in the limit $\frac{N_o\Delta f}{_{sc}}\to \infty$, the \gls{mmse} expression converges to the hermitian (matched filter) which maximizes the received SNR. Thus, \gls{mmse} offers a balance between interference suppression and noise reduction. 


Finally, the final \textit{effective} transmit \gls{bf} vectors at the \gls{bs} for \gls{dl} are obtained by first computing
$$\left(\tilde\vv_1^{\gls{mmse}}[k]\dots \tilde\vv_{N_u}^{\gls{mmse}}[k]\right)= \left(\vv_1^{CB}\dots \vv_{N_u}^{CB}\right)\V_{\gls{mmse}}[k],$$
and then introducing the following normalization to preserve the transmitted power constraint in each layer:
$$\vv_u^{\gls{mmse}}[k]=\tilde\vv_u^{\gls{mmse}}[k]/|\tilde\vv_u^{\gls{mmse}}[k]|.$$
Introducing these effective vectors into \eqref{eq:sinrdl}, instead of the auxiliary \gls{cbf} vectors we discussed earlier, results in the new \gls{sinr} values of the \gls{mmse} technique. For \gls{ul}, an equivalent hybrid combining at the BS receiver can be formulated by trasposing the matrices described in this section. 

\subsection{Scheduling}

We assume that the scheduler produces allocation decisions assigning transmissions to the \gls{tdma}+\gls{sdma} grid $(n,\ell)$ periodically \cite{TS38211v16}. We assume that the first and the last symbol of each \gls{nr} 14-symbol slot are reserved for the transmission of the \gls{pdcch} and \gls{pucch}, respectively. Data transmissions are assigned to the symbols 2 to 13 of the slot, and all symbols are ``flexible," meaning that they can be employed for \gls{dl} or \gls{ul} at the scheduler's decision \cite{TS38211v16}. We assume perfect channel estimation and do not model \glspl{dmrs} explicitly. We assume that the minimum data allocation unit is 1 \gls{ofdm} symbol of data transmission.

\begin{figure}
    \centering
    \subfigure[At $T_1$ allocation $\#1$ does not observe \gls{dmrs} from other layers and cannot design a \gls{mmse} precoding. When Allocation $\#2$ starts at $T_2$, it does not observe other \glspl{dmrs} either. After $T_2$, both allocations experience interference as in \gls{cbf} even though \gls{smbf} is supported.]{
        \includegraphics[width=.7\columnwidth]{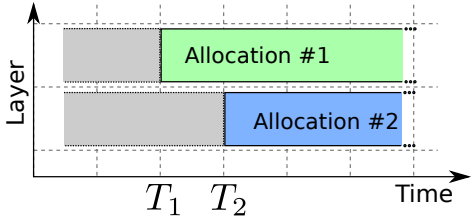}
        \label{fig:bfschedasync}
    }
    \hspace{.1in}
    \subfigure[After the transmission in the top layer ends at $T_1$, the scheduler leaves a padding symbol without signal and Allocation $\#1$ starts at $T_2$. Both layers start a new allocation simultaneously, they observe the other's \gls{dmrs} and implement \gls{smbf} successfully. However, the resource region $T_2-T_1$ in the top layer is wasted. ]{
        \includegraphics[width=.7\columnwidth]{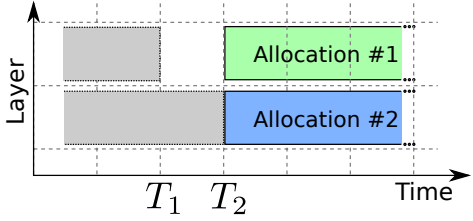}
        \label{fig:bfschedsync}
    }
    \caption{\gls{smbf} conflict for parallel allocations with different start time.}
    \label{fig:bfschedconflict}
    \vspace{-0.4cm}
\end{figure}

Since each allocation has only one front-loaded \gls{dmrs}, the \gls{bf} vector selected at the start of the transmission may not change until the allocation ends. This means that, when two transmissions overlap with different starting instants, they may not observe each other's references \gls{dmrs} and employ \gls{smbf} to reduce their mutual interference (Fig. \ref{fig:bfschedasync}). For this reason we design a scheduler that introduces blank symbols at the end of some transmissions and enforces equal start times for all simultaneous transmissions of different layers (Fig. \ref{fig:bfschedsync}).

\begin{figure}
    \centering
        \includegraphics[width=.8\columnwidth]{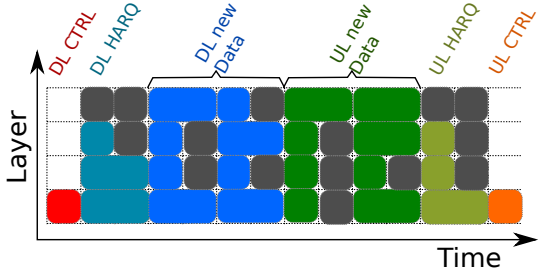}
    \caption{\gls{pmrs} example. The resources in gray are wasted as padding to guarantee equal start times for all allocations.}
    \label{fig:schedexample}
\end{figure}

Given $N_\ell$ layers, $N_s$ symbols and $N_u$ total UEs, our \gls{pmrs} equally divides the subframe in $N_b=\lceil N_u/N_\ell \rceil$ ``\gls{sdma} bundles." Each bundle consists in $N_\ell$ concurrent transmissions with the same start time allocated to different layers. Bundles are further time-multiplexed over the full subframe, where each bundle duration in time is exactly $N_a=\lfloor N_s/N_b \rfloor$ symbols. Layers that have fewer symbols to transmit start at the same time as the rest of their bundle, but end transmitting sooner. The interval between the end of any transmission and the start of the next bundle is a ``padding" of wasted symbols (Fig. \ref{fig:schedexample}). Within each bundle and for each layer, a different UE is selected. If $N_u>N_s\times N_\ell$, then some UEs are left unserved and become the first UEs in the list for the next subframe in a \gls{rr} fashion.

\section{Performance Evaluation}
\label{sec:perf_eval}

We implemented an \gls{mumimo} \gls{hbf} extension for the ns-3 \gls{mmwave} module introduced in \cite{mezzavilla2018end}. Besides the implementation of the \gls{hbf} and \gls{mumimo} features, we made adjustments for a more realistic simulation of \gls{5g} networks. Instead of the NYU channel model \cite{mezzavilla2018end}, we adopt the \gls{3gpp} channel model \cite{zugno2020implementation}. In addition, the \gls{ofdm} resource grid parameters (bandwidth, subcarrier spacing, symbol duration, and number of slots per frame) reflect those of \gls{nr}, as described in Sec.~\ref{sec:system} and~\cite{TS38211v16}. Notice that the ns-3 mmWave module assumes that control signaling is ideal and messages are never lost or corrupted.

In our implementation we have introduced modifications to numerous C++ classes in the ns-3 mmwave module.
Notably, the antenna array module now supports multiple antenna ports, with different \gls{bf} configurations. Moreover, the 3GPP channel model implementation has been extended to account for the multi-layer interference of Eq.~\eqref{eq:sinrdl} and Eq.~\eqref{eq:sinrul}, while the channel abstraction code and the physical layer implementation have been refactored to support multiple \gls{sdma} asynchronous layers (i.e., transmissions from a single entity). The \gls{bf} strategies described in Sec.~\ref{sec:system} have been implemented in a plug-and-play fashion, leveraging a novel, flexible \gls{bf} module. Finally, we updated the ns-3 mmwave module \gls{mac} layer to support multiple asynchronous layers, by properly accounting for the mapping of upper layer PDUs to mmwave Transport Blocks on different antenna ports, the management of \gls{harq} retransmissions, the CQI estimation, and the control signaling. The \gls{mac} layer also features a plug-and-play implementation of the scheduler introduced in Sec.~\ref{sec:system}, which is backward compatible and allows comparison with the other scheduling strategies implemented in the ns-3 mmWave module~\cite{mezzavilla2018end}. We refer the reader to the publicly available Github repository with the \gls{hbf} extension for additional details.

\subsection{Simulation Scenario}
\label{sec:scenario}


We consider a random \gls{mmwave} cellular system with one BS located at the origin of the coordinates (0,0) with a height of $25$~m, and 7 UEs located at random positions uniformly distributed in a disc of radius $100$~m with a height of $1.6$~m. We generate and average the results over 20 such random deployments, UE locations and channels. Due to the considerable pathloss in \gls{mmwave}, we assume inter-cell interference is severely attenuated and it is sufficient to simulate one cell.

We configured the \gls{nr} \gls{ofdm} waveform with numerology $\mu=2$, which corresponds to a subcarrier spacing of $60$~kHz. The central frequency is $28$~GHz and the bandwidth $198$~MHz is divided into 275 \glspl{rb}, each including 12 subcarriers. There are 4 slots per subframe with duration 250~$\mu$s, and the \gls{ofdm} symbol duration is $17.85$ $\mu$s including the CPs. We adopt the channel model described in \gls{3gpp} TR~38.901~\cite{TR38901} and consider the ``Urban Macro'' scenario. The transmission power is 30 dBm, and the receiver noise figure 5 dB. We consider a $8\times 8$ \gls{upa} with either $1$ or $4$ layers in the \gls{bs} and a $4\times 4$ \gls{upa} with $1$ layer for the \glspl{ue}.

\subsection{Comparison of \gls{bf} Solutions}
\label{sec:bfcomparison}

We compare the \gls{cbf} scheme that focuses on improving the \gls{snr} and is sufficient in single-layer cases versus our proposed \gls{smbf} scheme. We use \gls{rlc} \gls{um} (i.e., without \gls{rlc} retransmissions), disable the \gls{harq} retransmissions, and use low-traffic application in the UEs. This makes it possible to probe the channel and \gls{bf} scheme at a constant rate, and to measure the statistics of the \gls{sinr} and \gls{bler} in the physical layer without disruptions by the upper layers.

The low rate application is a constant traffic generator that produces a packet of 1500 bytes every 1500 $\mu$s in each \gls{ue}. Roughly speaking, when the \gls{mcs} coding rate is greater than 3.64 bits per subcarrier, the 3300 subcarriers can carry a full packet in a single \gls{ofdm} symbol. This means that the scheduler receives a demand for 14 symbols in one out of every six slot of $250$ $\mu$s. The frame has thus plenty of \glspl{rb} to satisfy the demand and the cell is lightly loaded.

\begin{figure}[t]
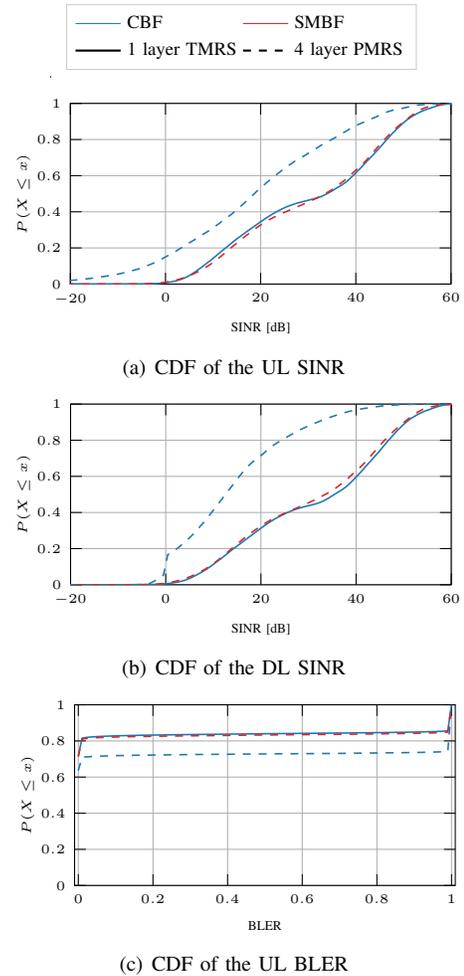

    \centering
      \setlength\fwidth{.75\columnwidth}
      \setlength\fheight{0.2\columnwidth}
\begin{tikzpicture}

\definecolor{color1}{rgb}{1,0.498039215686275,0.0549019607843137}
\definecolor{color2}{rgb}{0.172549019607843,0.627450980392157,0.172549019607843}
\definecolor{color4}{rgb}{0.580392156862745,0.403921568627451,0.741176470588235}
\definecolor{color5}{rgb}{0.549019607843137,0.337254901960784,0.294117647058824}
\definecolor{color3}{rgb}{0.83921568627451,0.152941176470588,0.156862745098039}
\definecolor{color0}{rgb}{0.12156862745098,0.466666666666667,0.705882352941177}

\begin{axis}[
legend cell align={left},
legend style={font=\scriptsize,at={(0.5,0.5)}, anchor=south, draw=white!80.0!black},
tick align=inside,
tick pos=left,
x grid style={white!69.01960784313725!black},
xmajorgrids,
xmin=0, xmax=840,
xtick style={color=black},
y grid style={white!69.01960784313725!black},
ymajorgrids,
ymin=0, ymax=810,
ytick style={color=black},
hide y axis,
hide x axis,
legend columns=2
]


 \addplot [solid, color0]
 table [row sep=crcr] {%
 -1 -1\\
 -2 -2\\
 };
 \addlegendentry{\gls{cbf}}


 \addplot [solid, color3]
 table [row sep=crcr] {%
 -1 -1\\
 -2 -2\\
 };
 \addlegendentry{\gls{smbf}}

\addplot [solid, black, thick]
table [row sep=crcr] {%
-1 -1\\
-2 -2\\
};
\addlegendentry{1 layer \gls{tmrs}}

\addplot [dashed, black, thick]
table [row sep=crcr] {%
-1 -1\\
-2 -2\\
};
\addlegendentry{4 layer \gls{pmrs}}
\end{axis}

\end{tikzpicture}
    \\
    \subfigure[\gls{cdf} of the \gls{ul} \gls{sinr}]{
      \setlength\fwidth{0.75\columnwidth}
      \setlength\fheight{0.45\columnwidth}
      \input{./figures/cdf_sinr_ul_rlcAm=False_interPacketInterval=1500_harq=False.tex}
    \label{fig:bfulsinr}
    }
    \subfigure[\gls{cdf} of the \gls{dl} \gls{sinr}]{
      \setlength\fwidth{0.75\columnwidth}
      \setlength\fheight{0.45\columnwidth}
      \input{./figures/cdf_sinr_dl_rlcAm=False_interPacketInterval=1500_harq=False.tex}
    \label{fig:bfdlsinr}
    }
    \subfigure[\gls{cdf} of the \gls{ul} \gls{bler}]{
      \setlength\fwidth{0.75\columnwidth}
      \setlength\fheight{0.45\columnwidth}
      \input{./figures/cdf_bler_ul_rlcAm=False_interPacketInterval=1500_harq=False.tex}
    \label{fig:bfulbler}
    }
    \caption{Comparison of the different \gls{bf} schemes.}
    \vspace{-0.4cm}
\end{figure}

We represent the received \gls{ul} \gls{sinr} \gls{cdf} for all transmission allocations in the simulation in Fig. \ref{fig:bfulsinr}. We compare 1 layer (solid) and 4 layer (dashed) cases. For the 1 layer case, we use the \gls{tmrs} without \gls{sdma} capabilities that was implemented in the previous versions of the ns-3 \gls{mmwave} module, with \gls{cbf}. For the 4 layer case we consider the \gls{pmrs} with both \gls{cbf} and \gls{smbf}. Since there is no self-interference, in the 1-layer case the \gls{sinr} is the same as the \gls{snr}. Moreover, with 4 layers, adopting a single-layer \gls{bf} scheme (\gls{cbf}) leads to \gls{sinr} degradation. Finally, for the \gls{smbf} scheme the \gls{sinr} \gls{cdf} is nearly identical to that of the single-layer \gls{cbf} case. This suggests that with \gls{smbf}, the \gls{sinr} is almost equal to the \gls{snr} and the scheme removes almost all inter-beam interference.

We represent the received \gls{dl} \gls{sinr} \gls{cdf} in Fig. \ref{fig:bfdlsinr}. The main difference with the \gls{ul} case is that in \gls{dl} the desired and interfering signals at each UE experience the same pathloss, and -20 dB \gls{sinr} outages with 4-layer \gls{cbf} rarely happen. On the other hand, the gap between the higher range of \glspl{sinr} achieved with 4-layer \gls{cbf} and 4-layer \gls{mmse} \gls{bf} schemes is wider than in \gls{ul}.

Finally we depict the instantaneous \gls{bler} \gls{cdf} for all \gls{ul} transmissions in Fig. \ref{fig:bfulbler}. The instantaneous \gls{bler} is dominated by outages when the channel has changed and the \gls{cqi} is outdated, as most transmissions experience either \gls{bler}$\leq 10^{-2}$ or \gls{bler}$=1$. As we can see, 4-layer \gls{cbf} has a much larger outage probability (lower step in the \gls{bler} \gls{cdf}) and results in more severe average \gls{bler} in the system. Again, \gls{smbf} behaves almost as a 1-layer \gls{cbf} situation. We do not depict the \gls{dl} \gls{bler} \gls{cdf} due to space constraints, as its insights were identical.

\subsection{Application Performance on Loaded Cell}
\label{sec:bf-sched-results}

Next, we consider the \gls{bler} and throughput with high traffic load applications. Qualitatively we wish to check that the \gls{pmrs} does not introduce too much padding so as to completely cancel the increased capacity of the multi-layer \gls{sdma} capabilities of the physical layer. Once again we use \gls{rlc} \gls{um} and disable the \gls{harq} retransmissions. For space constraints we leave the impact of retransmissions in application throughput and delay for future extensions of our work. 

Since we have already determined the best \gls{bf} scheme for each number of layers, we consider the \gls{tmrs} 1-layer scheduler with \gls{cbf}, and our proposed \gls{pmrs} for 4 layers with \gls{smbf}. 

The high-rate application is a constant bit rate source that generates a packet of 1500 bytes every 150 $\mu$s in each \gls{ue}, with a symmetric traffic in uplink and downlink. For every slot of $250$ $\mu$s, the scheduler always receives requests for at least $\sim$~23 symbols. In the 1-layer case there are $12$ data symbols per slot, which are not enough to allocate all the demand. In the 4-layer case, there are $12\times4$ available data symbols, i.e., more than enough when the users demand $\sim$~23 symbols.

\begin{figure}[t]
    \centering
    \subfigure[Average \gls{bler}]{
      \setlength\fwidth{0.75\columnwidth}
      \setlength\fheight{0.5\columnwidth}
\begin{tikzpicture}

  \definecolor{color1}{rgb}{1,0.498039215686275,0.0549019607843137}
  \definecolor{color2}{rgb}{0.172549019607843,0.627450980392157,0.172549019607843}
  \definecolor{color4}{rgb}{0.580392156862745,0.403921568627451,0.741176470588235}
  \definecolor{color5}{rgb}{0.549019607843137,0.337254901960784,0.294117647058824}
  \definecolor{color3}{rgb}{0.83921568627451,0.152941176470588,0.156862745098039}
  \definecolor{color0}{rgb}{0.12156862745098,0.466666666666667,0.705882352941177}

\begin{axis}[
  ybar,
tick align=inside,
tick pos=both,
x grid style={white!69.01960784313725!black},
xtick style={color=black},
xtick={0,1,2,3},
xticklabels={\gls{tmrs}-1L, \gls{pmrs}-4L},
y grid style={white!69.01960784313725!black},
ylabel={BLER},
ymin=0, ymax=0.5,
ytick style={color=black},
ytick={0,0.1,0.2,0.3,0.4,0.5},
yticklabels={0.0,0.1,0.2,0.3,0.4,0.5},
bar width=30pt,
xmin=0, xmax=1,
enlarge x limits=0.45,
legend columns=2, 
legend style={font=\tiny,at={(0.02,0.98)}, anchor=north west, draw=white!80.0!black},
label style={font=\tiny},
tick label style={font=\tiny} 
]

\addlegendimage{ybar,ybar legend, color0, fill=color0,fill opacity=0.4};
\addlegendentry{DL}

\addlegendimage{ybar,ybar legend, color1, fill=color1,fill opacity=0.4, postaction={pattern=north east lines}};
\addlegendentry{UL}

\addplot[color0, fill=color0, fill opacity=0.4] coordinates {(0,0.123515226065967) (1, 0.159321238785321) };

\addplot[color1, fill=color1, fill opacity=0.4, postaction={pattern=north east lines}] coordinates {(0,0.242682975360048) (1,0.136730253662689)};

%

%
\end{axis}
\end{tikzpicture}
    \label{fig:schedbler}
    }
    \hspace{.01in}
    \subfigure[Throughput]{
      \setlength\fwidth{0.75\columnwidth}
      \setlength\fheight{0.5\columnwidth}
\begin{tikzpicture}

  \definecolor{color1}{rgb}{1,0.498039215686275,0.0549019607843137}
  \definecolor{color2}{rgb}{0.172549019607843,0.627450980392157,0.172549019607843}
  \definecolor{color4}{rgb}{0.580392156862745,0.403921568627451,0.741176470588235}
  \definecolor{color5}{rgb}{0.549019607843137,0.337254901960784,0.294117647058824}
  \definecolor{color3}{rgb}{0.83921568627451,0.152941176470588,0.156862745098039}
  \definecolor{color0}{rgb}{0.12156862745098,0.466666666666667,0.705882352941177}

  \begin{axis}[
    ybar,
  tick align=inside,
  tick pos=both,
  x grid style={white!69.01960784313725!black},
  xtick style={color=black},
  xtick={0,1,2,3},
xticklabels={TMRS-1L, PMRS-4L},
  y grid style={white!69.01960784313725!black},
  ylabel={Throughput [Mbps]},
  ymin=0, ymax=60e1,
  ytick style={color=black},
  bar width=30pt,
  xmin=0, xmax=1,
  enlarge x limits=0.45,
  legend columns=2,
  legend style={font=\tiny,at={(0.02,0.98)}, anchor=north west, draw=white!80.0!black},
    label style={font=\tiny},
    tick label style={font=\tiny} 
  ]

  \addlegendimage{ybar,ybar legend, color0, fill=color0,fill opacity=0.4};
  \addlegendentry{DL}

  \addlegendimage{ybar,ybar legend, color1, fill=color1,fill opacity=0.4, postaction={pattern=north east lines}};
  \addlegendentry{UL}

  \addplot[color0, fill=color0, fill opacity=0.4] coordinates {(0,334.253935) 
  (1,424.865305) };
  
  \addplot[color1, fill=color1, fill opacity=0.4, postaction={pattern=north east lines}] coordinates {(0,182.657610) 
  (1,459.142755)};

  %
  
\end{axis}

\end{tikzpicture}
    \label{fig:schedulDataRx}
    }
    \caption{Comparison of the different scheduling strategies.}
    \vspace{-0.4cm}
\end{figure}
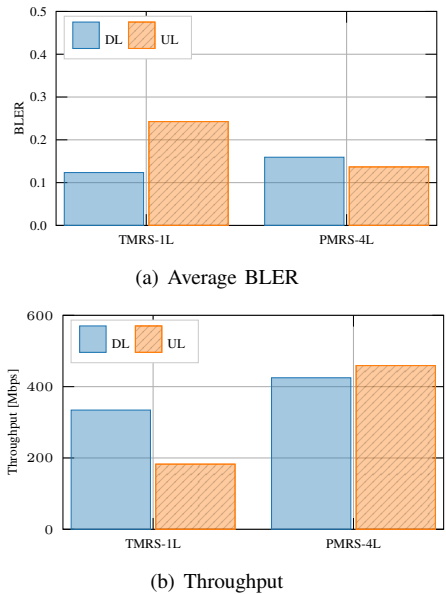

Figure \ref{fig:schedbler} reports the average \gls{dl} and \gls{ul} \gls{bler} for the two cases. The \gls{bler} of \gls{pmrs} with 4 layers is comparable to that of \gls{tmrs} with 1 layer. Recalling our prior remark that \gls{bler} is mostly driven by \gls{cqi} outage, this is consistent with the \gls{sinr} plots discussed in the previous section. Since \gls{smbf} removed almost all interference, both schemes experienced comparable \gls{cqi} and outage events.

Figure \ref{fig:schedulDataRx} depicts the average throughput in the simulations. The sum source rate is  $560$~Mbps. For \gls{tmrs} with \gls{cbf}, we see that the throughput is 330~Mbps in \gls{dl} and 180~Mbps in \gls{ul}, with significant asymmetry and lower value than the offered traffic. This is because the demand exceeds the number of data symbols of the 1-layer frame even with the maximum \gls{mcs} rate. In the 4-layer case, the cell is not saturated, and the throughput with \gls{pmrs} exceeds 420~Mbps in \gls{dl} and 450~Mbps in \gls{ul}. This shows the main advantage of \gls{sdma} \gls{mumimo} in \gls{mmwave} networks, i.e., \textit{an increase in the number of available \glspl{rb} by a factor of $N_{layers}$ allows the network to support much more traffic.} Particularly in our simulation a delivered traffic that is $2\times$ the capacity of the single-layer frame was achieved without stressing the system. The \gls{sdma} scheme could still saturate for even higher application rates, or experience issues with delay or with retransmissions either by \gls{harq} or \gls{tcp}, which we leave for future extensions.

\section{Conclusions}
\label{sec:conclusions} 

We have studied \gls{mumimo} \gls{hbf} implementations for \gls{3gpp} \gls{nr} \gls{mmwave} end-to-end cellular systems. We have shown that \gls{sdma} greatly increases the system capacity. Moreover, by associating each frequency-flat \gls{bf} vector to a separate antenna port, the \gls{mumimo} signal processing can be handled in a space of reduced dimensions instead of the large arrays characteristic of \gls{mmwave}. It is necessary to alleviate the inter-beam interference in order to improve the \gls{sinr}, as otherwise if we merely used separate analog beams the \gls{sinr} would degrade significantly. We reveal a conflict between the design of \gls{mumimo} schedulers and \gls{mumimo} \gls{bf}, which stems from the unique front-loaded  \gls{dmrs} per transmission. In future work we intend to study feedback overhead reduction, asynchronous scheduling without padding, and the effect of retransmission schemes on throughput and the delay observed by the applcations. We also leave for future work the study of non-linear \gls{mumimo} signal processing methods, such as joint sphere decoding or successive interference cancelling, which can theoretically outperform \gls{mmse} in terms of \gls{bler}.

\bibliographystyle{IEEEtran}
\bibliography{bibl.bib}

\end{document}